\documentclass[a4paper,11pt]{article}
\usepackage{pos}
\usepackage[style=numeric,sorting=none]{biblatex}
\usepackage{subcaption}
\usepackage{graphicx}
\smallskipamount

\title{The importance of being discrete }

\author*{Goran Senjanovi\'c}

\affiliation{International Centre for Theoretical Physics,\\
Trieste, Italy}

\emailAdd{goransenjanovic@gmail.com}

\abstract{Although continuous symmetries may be more appealing, especially the local gauge ones, I argue that discrete symmetries may still play a fundamental role in shaping our understanding of the physics beyond the Standard Model. 
 I exemplify this on the case of left-right symmetry and recall how its spontaneous breaking leads naturally to new weak interactions, and even more interestingly, to neutrino mass and the seesaw mechanism behind its smallness. The minimal realization of this program is shown to be a self-contained theory of the origin and nature of neutrino mass, and of quark mixing angles. I also discuss some new results on the extension of the theory based on Pati-Salam quark-lepton symmetry, and its embedding in $SO(10)$ grand unified theory, where LR symmetry becomes gauged discrete transformation in the form of charge conjugation.
}

\addtocontents{toc}{\setcounter{tocdepth}{4}} 
\FullConference{%
  DISCRETE2024, 9th Symposium on Prospects in the Physics of Discrete Symmetries\\
  December 2 - 6, 2025\\
  Ljubljana, Slovenia}

\tableofcontents

\begin{document}
\maketitle
\section{Introduction}

Continuous symmetries play a profound role in physics, in particular local gauge symmetries since they  require the existence of mediators of forces - long range if a symmetry is unbroken, and short range when it is broken. And what makes it even more important in the case of non-Abelian symmetries, is that they can be only spontaneously broken in order that a relevant theory be renormalisable. 

While less attractive and not directly related to any dynamics,  global continuous symmetries are still of great use, since they can provide  bookkeeping. Nice examples are provided by the global $SU(2)$ isospin symmetry, or baryon and lepton number $U(1)$ accidental global symmetries of the Standard Model. They need not be exact to be useful, and they can be equally explicitly as spontaneously broken. 

If broken spontaneously, global continuous symmetries lead to the existence of massless scalars, the Goldstone bosons, whose dynamics is rather appealing - even if a symmetry in question is not exact. In this case, if explicit breaking is small, we end up with pseudo-Goldstone bosons, which although not massless, are typically much lighter that the physical scale in question. A beautiful example is provided by the pions, the pseudo-Goldstone bosons of the chiral $SU(2)\times SU(2)$ symmetry of massless up and down quarks. 

What about discrete symmetries? At first glance, they would appear superfluous, deprived of any interesting dynamics. Appearances may be deceptive, though. Particularly appealing examples are provided by the symmetries once held sacred, but turned out broken by weak interaction: parity $P$ and its product with charge conjugation $C$, namely the $CP$ symmetry. The former is broken maximally by the charged weak interaction, while the latter one barely. The maximal breakdown of parity is at the core of the SM - the simplicity of the spontaneous symmetry breaking provided by the single Higgs doublet, works precisely due to the maximal parity breaking, with Left-Handed (LH) fermions being the weak doublets, and their Right-Handed (RH) counterparts the weak singlets. 

And yet, the fathers of the idea of parity breaking, Lee and Yang, in their classic paper~\cite{Lee:1956qn} suggested that deep down, at the fundamental level, parity could be restored. After all, the LR symmetry cannot be more natural, it is one of first symmetries that a child sees.  When the experiment showed that parity is broken, and broken maximally, this was forgotten by basically everybody, including the authors~\footnote{T.D.Lee, private communication, Beijing 2007.} This idea would be revived years later~\cite{Pati:1974yy,Mohapatra:1974gc,Senjanovic:1975rk,Senjanovic:1978ev} in the context of modern gauge theories, when it emerged that such theories are indeed renormalisable. By postulating that electroweak interactions are LR symmetric, the theory immediately predicted the existence of new weak gauge bosons, the RH analogs of the SM ones. Equally, if not more, important, the theory predicted neutrino mass decades before it would be confirmed experimentally. These predictions stemmed directly from the minimal gauge group structure $G_{LR} = SU(2)_L \times SU(2)_R \times U(1)_{B-L} $. 

Detailed, quantitative predictions depend, however,
on the choice for the Higgs sector responsible for the breaking of the original gauge group $G_{LR}$ down to the Standard Model gauge group 
$G_{SM} = SU(2)_L \times U(1)_Y$. There are two natural minimal choices, either left and right 
doublets~\cite{Pati:1974yy,Mohapatra:1974gc,Senjanovic:1975rk,Senjanovic:1978ev}, or 
triplets~\cite{Minkowski:1977sc,Mohapatra:1979ia,Mohapatra:1980yp}. In the former case, one is led to the Dirac neutrino mass scenario, while in the latter one, one ends up with Majorana neutrino masses. The Majorana case has won over the years, since it provides a rationale for the smallness of neutrino mass through the seesaw mechanism~\cite{Minkowski:1977sc,Mohapatra:1979ia}. I should stress that the seesaw mechanism does not require~\cite{Yanagida:1979as} LR symmetry, but LR makes it simply more natural - even if it is sometimes hidden in the gauge symmetry 
itself~\cite{GellMann:1980vs,Glashow:1979nm}, as in the case or $SO(10)$ grand unified 
theory~\cite{Fritzsch:1974nn,Georgi:1974my}. It is LR symmetry which requires the existence of the RH neutrino - otherwise postulaed {\it ad hoc} in the seesaw mechanism based on adding RH neutrino to the SM - that then triggers small neutrino mass. 

It took almost forty years to realise that this minimal LR symmetric model based on the seesaw mechanism is actually a self-contained theory of neutrino mass, in complete analogy with the SM for the masses of charged fermions - there are a plethora of new processes predicted by the knowledge of neutrino mass, just as the knowledge of the charged fermion mass determines the Higgs decay into such fermion anti-fermion pairs. 
One should keep in mind, though, that the Dirac picture is still perfectly valid and should not be discarded {\it apriori}. Although it is hard to understand in this case why neutrino is so much lighter than the electron, the Dirac scenario has interesting predictions for the RH leptonic mixings~\cite{dartagnan}. Moreover, it predicts a ratio of the new neutral versus charged gauge boson mass (different from the Majorana case), and can thus be tested at hadron colliders. 

However, we will review here the Majorana picture, which offers a much richer dynamics,
in particular, a number of new lepton number violating processes. Two such processes stand out, the neutrinoless double beta decay ($0 \nu 2 \beta$) at low energies, and the so-called Keung-Senjanovic (KS) process at high energies. The latter one is specially interesting since it allows to  probe the Majorana nature of the heavy RH neutrino mass. Moreover, within the minimal model, these two processes are deeply related, and the KS process allows one to make predictions for the neutrinoless double beta decay~\cite{Tello:2010am}.

This is the main reason behind my focusing here on parity, and not on $CP$. After all, the breaking of  $CP$ plays no fundamental role in the Standard Model construction, it just requires the Yukawa couplings to be complex. It is interesting, though, that the idea to break  $CP$ spontaneously arised before the analogous one for $P$, but as such does not lead to new exciting dynamics - it simply postulates Yukawa couplings to be real, by doubling the Higgs sector. Therefore, in the rest of this short review, I will discuss only the consequences of the spontaneous breaking of parity symmetry. Strictly speaking, one can swap $P$ for $C$, since they are equally LR symmetric transformations. In fact, in the $SO(10)$
grand unified theory, the LR symmetry is a finite gauge transformation in the form of charge conjugation.

The rest of my talk is then organised as follows. In the next section I discuss the minimal LR symmetric model that leads to neutrino Majorana picture. This is the core of the talk, and for all practical purposes the reader need read beyond it. Next, the section \ref{PS} is devoted to the generalisation of this model to the Pati-Salam theory, which includes on top the idea of quark-lepton unification. This theory unfortunately requires the scale of new physics to be astronomically large, but offers beautiful physics of magnetic monopoles. In the 
 section \ref{so10} I focus on the embedding of the Pati-Salam theory into the  $SO(10)$ grand unified theory~\cite{Fritzsch:1974nn}, which moreover unifies a family of fermions into a single spinorial representation, augmented by the right-handed neutrino. If the reader has doubts whether the LR symmetry is really called for, the $SO(10)$ theory is then tailor made for her, since it is automatiucally LR symmetric - namely, charge conjugation is a discrete gauge symmetry. Section \ref{summary} is finally devoted to our conclusions and outlook.



\section{Minimal LR symmetric theory}
\label{LR}

In this section I briefly review the salient features of a minimal LR symmetric model (LRSM) based on the neutrino Majorana picture. For more details, the interested reader is referred  
to~\cite{Tello:2012qda,Senjanovic:2023czt}, while the reader in need of truly pedagogical {\it expose} should consult~\cite{Melfo:2021wry}.
A minimal LR symmetric model (LRSM) is based on the gauge group 
\begin{equation}
	\label{eq:glr}
	G_{LR}= SU(2)_L \times SU(2)_R \times U(1)_{B-L}\,,
\end{equation}
augmented by the LR symmetry, here chosen to be parity $P$. The electromagnetic charge is thus given by 
\begin{equation}
	\label{eq:qem}
	Q_{em}= T_{3L} + T_{3R} + \frac{B-L}{2}\,.
\end{equation}
The $U(1)$ generator, unlike the SM hypercharge, is a physical number, the difference between baryon and lepton numbers, corresponding to the accidental anomaly-free global symmetry of the SM.
Since it is gauged here, anomaly cancellation now requires the existence of the Right-Handed (RH) neutrino. Needless to say, so does the LR symmetry itself, which dictates the fermion assignment 
  \begin{equation}\label{lrferm}
  \left(\begin{array}{c} u \\ d \end{array} \right)_L, \,\,\, \left(\begin{array}{c} \nu \\ e \end{array} \right)_L\,\,\,\,\,\,\,\,\,;\,\,\,\,\,\,\,\,\, \left(\begin{array}{c} u \\ d \end{array} \right)_R, \,\,\, \left(\begin{array}{c} \nu \\ e \end{array} \right)_R\,.
 \end{equation}
Clearly, the gauge structure forces neutrinos to be massive - independently of the choice of the Higgs sector - just like their siblings, the charged fermions. The nature and the origin of neutrino mass, depends however on the Higgs multiplets used at the first stage of symmetre breaking, when $G_{LR}$ breaks down to $G_{SM}$. In this theory, this is achieved by $SU(2)_{L,R}$
triplets $\Delta_{L,R}$, carrying $B-L = 2$. It is easily shown that for a range a parameters in the Higgs potential, the symmetry breaking pattern takes the form
  \begin{equation} \label{LRvevs}
  \langle \Delta_L \rangle = 0 \,\,; \,\,\, \, \,\,\,\langle \Delta_R \rangle \neq 0,
   \end{equation}
In turn, one obtains the relation for the masses of the new heavy charged and neutral gauge bosons 
  \begin{equation} \label{gaugemasses}
  M_{Z_R} \simeq \sqrt 3 M_{W_R} \propto \langle \Delta_R \rangle,
   \end{equation}
implying a clear way or ruling out the theory by finding the neutral gauge boson $Z_R$ before the charged $W_R$. It should also be stressed that the relation would change in the case of the Higgs triplets being swapped by doublets, making $Z_R$ lighter by a $\sqrt 2$. As in the case of the SM, the mass relation between the gauge bosons distinguishes between different choices for the Higgs boson sector. 

\subsection{Leptonic sector} 
\label{leptonsector}

We turn our attention first to the lepton sector, since this is where LR symmetry plays a crucial role in providing small neutrino mass and leading to Lepton Number Violation (LNV) both at low and high energies.

\paragraph*{Seesaw mechanism and how to untagle it}

Moreover - this is a central result - the RH neutrino gets a Majorana mass $M_N \propto \langle \Delta_R \rangle$ and becomes a heavy neutral lepton $N$, potentially accessible at the LHC through the interactions with $W_R$. In turn, via the Dirac mass term $M_D$ with the usual LH neutrino, it forces neutrino to pick a small Majorana mass 
  \begin{equation}
     M_\nu   \simeq   - M_D^T  \frac{1}{M_N} M_D\,. 
     \label{seesaw}
     \end{equation}
This is the celebrated seesaw mechanism, that makes neutrino naturally light compared to the charged leptons. Even more important, it connects neutrino mass to the amount of parity violation in nature, since $M_N \propto M_{W_R}$: Neutrino is light since parity is (almost) maximally broken in beta decay. The SM result of massless neutrinos is recovered in the limit of infinitely heavy $W_R$, as physically expected. And RH neutrino in this theory has new gauge interactions that allow it be produced at hadron collider - it is not a phantom particle you integrate out just to make neutrino massive, and light.

There is more to it, though. The LR symmetry forces the Dirac mass matrix to be either (almost) hermitian in the case of parity, or symmetric in the case of charge conjugation. While in the former case, the analysis is more involved~\cite{Senjanovic:2016vxw,Senjanovic:2018xtu,Senjanovic:2019moe}, the end result is the same: The seesaw mechanism can be untangled. I illustrate it here for the case of $C$~\cite{Nemevsek:2012iq}
 \begin{equation}\label{untangle}
 M_D^{LR} = i M_N \sqrt  {M_N^{-1} M_\nu}.
 \end{equation}
In other words, one can determine $M_D$ by measuring $M_\nu$ at low energies  and $M_N$ at hadron colliders, once $N$ gets produced. Keep in mind, that LR symmetry is indispensable here - without it, $M_D$ is undetermined, due to an arbitrary complex orthogonal matrix $O$ spoiling the above relation~\cite{Casas:2001sr}
\begin{equation}\label{naive}
 M_D^{seesaw} = \sqrt{m_N}  O \sqrt{M_\nu},
  \end{equation}

Determination of $M_D$  predicts in turn the rate for the $N$ decay into the $W$ boson (if $N$ heavier than $W$) and the charged lepton $\ell$~\cite{Nemevsek:2012iq}
\begin{equation} \label{Ndecays}
\Gamma(N \to W \ell) \propto \frac{m_N^2}{M_W^2} \, m_\nu\,,
\end{equation}
where for the sake of illustration I assumed the same LH and RH leptonic mixing angles. This allows us to probe the Higgs origin of neutrino mass, just as the SM Higgs boson decay into electron and positron allows to probe the origin of electron mass. Notice that I did not specify the charges - since $N$ is a Majorana particle it decays equally into charged leptons and their anti particles $\Gamma(N \to W^+ \ell) = \Gamma(N \to W^- \bar \ell)$ . It should be stressed that there are a number of similar processes which serve as a way of testing the origin of neutrino mass~\cite{Senjanovic:2018xtu}. In short, the LRSM is truly a predictive theory of neutrino mass - all that is need is to produce $N$, and measure the mass matrix $M_N$, since we are close to determining $M_\nu$.

\paragraph*{LNV and how to probe the nature of neutrino mass} 

The core of the LR symmetric theory lies in the fact of the heavy RH neutrino having new gauge interactions and thus potentially reachable at the LHC or the next hadron collider. Due to its Majorana nature, once produced it would lead to  the production of same sign charged leptons accompanied by jets~\cite{Keung:1983uu}. This is the KS process, a generic feature of all models leading to Majorana masses, a paradigm for LNV at hadron colliders. 

There is more to it though. The KS process allows furthermore to test directly the Majorana nature of N, since once produced on-shell it must decay equally into charged leptons and anti-leptons. Moreover, it allows also to probe directly Lepton Flavor Violation at colliders and relate it to low energy such processes. For a comprehensive study of all of this, see~\cite{Tello:2012qda}.

\subsection{Quark sector} 
\label{quarksector} 

The main question in the quark sector is the form of the RH analogue of the CKM matrix. Since parity is almost maximally broken in charged weak interaction, one would naively think that the RH mixings are not related to the LH ones. However, the softness of spontaneous symmetry brwaking keeps the memory of the initial parity symmetry, and remarkably, the RH quark mixing matrix $V_R$ can be determined as a function of the LH mixing matrix $V_L$~\cite{Senjanovic:2014pva,Senjanovic:2015yea}
\begin{equation} \label{eq:master}
(V_R)_{ij} \simeq (V_L)_{ij} - i \epsilon \frac{(V_L)_{ik} (V_L^\dagger m_u V_L)_{kj}} {m_{d_k} + m_{d_j}}.
\end{equation}
where $\epsilon $ is a small parameter to be determined experimentally. It is an easy exercise to show that the LH and RH mixing angles are basically the same, an important result that allows to set directly limits on the $W_R$ mass from experiment. The main difference between the two sectors are new phases in $V_R$, but from the above expression can be predicted theoretically as a function of $\epsilon$. This is a remarkable result, which was decades was one of the main, if not the main, challenges for the theory - it was finally completed due to the dedicated efforts of Vladimir Tello. 

\subsection{Collider searches} 
\label{collider} 

While low energy processes can set lower limit on the scale of LR symmetry breaking, LHC has caught up with them some time ago. Both CMS and ATLAS have a dedicated search program for the physics of $W_R$, both through hadronic and leptonic decays. The crucial role in this is played by the knowledge of RH quark mixing  matrix $V_R$ in ~\eqref{eq:master}, which sets $W_R$ production in the quark-anti quark scatterings. 

The most reliable limit on $M_R $ comes from $W_R$ decays into quark jets, since they are free of any uncertainties, leading to 
$M_{W_R} \geq 4.6$TeV~\cite{ATLAS:2023ibb, CMS:2023gte}. There are also $W_R$ searches from the KS process, whose results depend on the mass of $N$ and are thus less direct - although comparable to the dijet decay ones - see~\cite{CMS:2021dzb,ATLAS:2023cjo}. A comprehensive study of hadron collider searches, with an emphasis on the KS process, has been performed in~\cite{Nemevsek:2018bbt}, where the authors discuss the reach of the LHC for the whole spectrum of the RH neutrino masses. A detailed study of LNV in various low and high energy processes can be found in ~\cite{Li:2024djp}, while  ~\cite{Kriewald:2024cgr} developed te program for precise predictions for the LRSM at colliders. Furthermore, the prospects for the next hadron collider are discusses in~\cite{Nemevsek:2023hwx}:  In other words, some half a century later the LR symmetric model has turned into the highly predictive analogue of the SM for the BSM physics. The way the SM allows us to probe the origin of charged fermion masses, on top of the gauge boson ones, the LRSM does it for the neutrino masses.

\subsection{Cosmological implications: domain walls, dark matter and leptogenesis} 
\label{dw} 

\paragraph*{Domain walls} 

It is well known that the spontaneous breaking of discrete symmetries leads to the existence of domain walls, in the form of kink solutions. These topological defects are a cosmological catastrophe~\cite{Zeldovich:1974uw} in the context of the standard cosmological model, as long as the scale of symmetry breaking is bigger than about 100 MeV. One is thus forced to get rid of them, before they dominate the energy content of the Universe.

Domain walls, as topological defects in general, are produced during a phase transition at high temperatures, via the so-called Kibble 
mechanism~\cite{Kibble:1976sj}. Thus, the simplest way out~\cite{Dvali:1995cc} of this impasse would be to keep symmetries broken at high temperature~\cite{Weinberg:1974hy,Mohapatra:1979qt,Mohapatra:1979vr,Mohapatra:1979bt}. However, in the minimal model the LR symmetry does get restored~\cite{Dvali:1996zr}. 

Another possibility, somewhat drastic, is to explicitly break discrete LR symmetry~\cite{Vilenkin:1984ib} - as long as such an effect is tiny, one can still speak of approximate spontaneously broken symmetry. Domain walls do get created in such a scenario, but decay before dominating the energy density of the Universe. One possible justification for such a violent act is the belief that quantum gravitational effects violate global symmetries, by analogy with the black hole physics. It is reassuring that even such miniscule effects, suppressed by powers of the Planck scale, do suffice to do the job~\cite{Rai:1992xw}, without spoiling any of the predictions of the theory.
An important outcome of these decaying LR domain walls are gravitational waves~\cite{Borah:2022wdy}, potentially detectable for a large $M_R$. 

\paragraph*{Dark matter}

A natural candidate for the dark matter of the universe is the lightest RH neutrino $N$, since if light enough, it is sufficiently long lived. This is an old idea~\cite{Dodelson:1993je} within the seesaw mechanism, and it fits nicely as the warm dark matter with mass in the keV region. This in turning forces the other two RH neutrinos to be light, with masses below GeV~\cite{Bezrukov:2009th,Nemevsek:2012cd} - and if this was realised in nature, one would lose the KS process as the direct LNV at colliders and a way of probing the Majorana nature of RH neutrinos.

\paragraph*{Leptogenesis}

The LRSM is tailor made for the leptogenesis scenario~\cite{Fukugita:1986hr}, where the out-of-equilibrium decays of the lightest RH neutrino $N$ create the lepton number of the universe, which is then transmitted to the baryon number through sphaleron interactions~\cite{Kuzmin:1985mm}. 

I should stress that this is a genuine theory of genesis, since the original baryon and lepton numbers of the universe get washed out at high temperatures above $m_N$. In other words, these numbers do not depend on the initial conditions. Due to $N$ interactions with RH gauge boson $W_R$, they stay in equilbirum unless $W_R$ is heavy enough $M_{W_R} \gtrsim 30$ TeV~\cite{Frere:2008ct}. If this was the mechanism of genesis, $W_R$ would not be accessible at the LHC, and for vested interests, I for one wish there was an alternative. I should add that this mechanism is in conflict with the dark matter being light $N$. In any case, if the KS process were to be observed at the LHC, we would lose both the darm matter and the leptogenesis mechanism within the minimal LR symmetric theory.
 
\subsection{LR symmetry and strong CP violation} 
\label{strongcp} 

In the SM the strong CP parameter 
$\bar \theta = \theta + arg det M_q$ is not related to any physical property of the theory, and is thus arbitrary. It is often claimed that its smallness poses a problem, but that is simply not correct. The only piece $arg det M_q$ that is affected by perturbation theory stays orders of magnitude below the experimental value $\bar \theta \lesssim 10^{-10}$, as long as the cutoff of the theory stays below Planck scale~\cite{Ellis:1978hq}. 

The situation is drastically different in the LRSM. One can even imagine that the theory predicts vanishing $\bar \theta$ due to the LR symmetry itself~\cite{Beg:1978mt,Mohapatra:1978fy}. This is a bit naive, since $\theta$ is not a hard parameter, and so it can break LR symmetry. However, the idea of LR symmetry requires that in the basis $\theta =0$, Yukawa couplings must be hermitian and therefore, the smallness of the weak contribution to the effective strong CP parameter, severely restricts symmetry breaking at the electroweak scale~\cite{Maiezza:2014ala}. This propagates even to the leptonic sector~\cite{Kuchimanchi:2014ota}, affecting the masses of the RH neutrinos~\cite{Senjanovic:2020int}. In a sense, strong CP violation, instead of being a problem, could be said to be a blessing, since it serves here to constrain the parameters of the theory, the same way the smallness of $K - \bar K$ mass difference restricts the charm quark mass in the SM~\cite{Glashow:1970gm}.

\section{Pati-Salam theory}
\label{PS}
One can marry LR symmetry with the quark-lepton unification, as suggested by Pati and Salam in their classic work~\cite{Pati:1974yy} that paved the road to grand unification. 
It is based on the symmetry group
\begin{equation}
G_{PS} \;=\; SU(4)_{ C}\,\times\,SU(2)_{ L}\,\times\,SU(2)_{ R} \times P\,,
\end{equation}
with $SU(4)_{ C}$ being the lepton addition to the usual $SU(3)_{ C}$ quark colour symmetry. 
 The minimal fermion content requires three copies of the following representations 
\begin{equation}
F_{ L} = (4_{ C},2_{ L},1_{ R}), \,\,\,
F_{ R} = (4_{ C},1_{L},2_{ R}). \,\,
\end{equation}
This encompasses all SM fermions, with the addition of the RH neutrino, as in the LRSM.

This is a beautiful theory, possessing many features of grand unification, including the existence~\cite{tHooft:1975psz} of magnetic monopoles~\cite{tHooft:1974kcl,Polyakov:1974ek}. 
This may come as a surprise to the reader since
naively one would imagine that monopoles require a single simple gauge group. The main difference between this theory and genuine grand unification is the absence of nucleon decay, which in principle allows for a possibility of fairly low unification scale, on the order of $10^5$ GeV~\cite{Dolan:2020doe}.

Besides magnetic monopoles, the theory possesses domain walls,  as the LR symmetric model. The way out of the notorious domain wall problem is simply to break LR explicitly as we said, in order to force the walls to decay. This way they can actually sweep the monopoles away, as suggested originally~\cite{Dvali:1997sa,Pogosian:2000xv} in the minimal $SU(5)$ grand unified theory~\cite{Georgi:1974sy}, thus solving two problems with one shot.

Recently, both the neutrino Dirac~\cite{Senjanovic:2025enc} and Majorana~\cite{Dvali:2023snt,Preda:2025afo} cases were studied in detail. In turns out that in the minimal versions, both cases lead to huge unification scale, above $10^{10}$ GeV or so, hopelessly out of direct experimental reach. Fortunately, the theory still leaves potentially observable phenomenological and cosmological imprints.

  \subsection{ Dirac neutrino case}
\label{diracnu}

We have recently revisited PS theory based on the Dirac neutrino picture~\cite{Senjanovic:2025enc}.  In this case, one has the following scalar multiplets:
\begin{equation}
\begin{split}
&H_{ L} = (4_{ C},2_{L},1_{ R})\,, 
\quad \,\,H_{ R} = (4_{ C},1_{ L},2_{ R})\, ,\\
&\,\Phi_1 = (1_{ C},2_{ L},2_{ R})\,,
\quad\Phi_{15} = (15_{ C},2_{ L},2_{ R})\,,
\end{split}
\end{equation}
    in obvious notation. The $H_{L,R}$ fields serve to break $G_{PS}$ down to $G_{SM}$, breaking parity in the process, while the $\Phi$ scalars serve to break the SM symmetry and provide masses to charged fermions.
At the renormalisable level, one needs both $\Phi_1$ and $\Phi_{15}$ in order to generate realistic mass spectra. Namely, $\Phi_1$ is a singlet under quark-lepton symmetry and, as such. could not distinguish between quark and lepton masses. 

The unification scale symmetry breaking proceeds along the lines discussed in the previous section in the LRSM: for a range of parameters of the Higgs potential, one has for the minimum~\cite{Senjanovic:1975rk,Senjanovic:1978ev}
\begin{equation}
 \langle H_L \rangle = 0 \,;
 \,\,\,\,\,\,\, \langle H_R \rangle \neq 0 \,,
\end{equation}
breaking both the PS gauge symmetry down the SM one, and in process also breaking the discrete LR symmetry. 
It can be easily seen that the SM scalar doublets drop from the unification constraints, and can thus be arbitrarily light. This has important cosmological implications. In particular, one of these doublets - the scalar analog of the weak lepton doublet, residing in $H_L$,  provides a natural dark matter candidate, a typical WIMP, with mass roughly below TeV. It can furthermore give rise to the strong first order phase transition, paving the way for the electroweak baryogenesis.  

  As we said, with small explicit breaking of LR symmetry gets rid of domain walls, which can then possibly sweep monopoles away. This beautiful proposal was first suggested in the context of the minimal $SU(5)$ grand unified theory, where one may have a weakly broken discrete symmetry with a small cubic term for the adjoint GUT Higgs field $24_H$. While there it is appealing, here it is a must since otherwise we end up with a domain wall catastrophe. In a sense, it may work only too well, leaving us without observable monopole fluxes, a sad outcome.  
  
  Fortunately, this mechanism leaves an important physical imprint in the form of gravitational waves. This was studied recently for the Dirac neutrino case, with the conclusion of gravitational waves having  their peak between 10 and 1000 Hz~\cite{Senjanovic:2025enc}, with potential reach in next generation searches. It would be interesting to see if and how it changes for the Majorana neutrino case.

 \subsection{ Majorana neutrino case}
\label{majorananu}

In this case, one swaps the $SU(2)_{L,R}$ doublets $H_{L,R}$ for the $SU(2)_{L,R}$, that couple to leptons and lead to the Majorana picture as in the LRSM discussed in the previous section.

First of all, it can be shown that the theory possesses scalars which masses can be small enough to induce neutrinoless double beta decay, with outgoing electrons being LH - and even dominate over the usual neutrino Majorana mass contribution~\cite{Dvali:2023snt}. Similarly, there are states that can induce observable $n - \bar n$ oscillations~\cite{Mohapatra:1980qe,Preda:2025afo}. Even more important, the scale of LR symmetry breaking - if one is willing to split it from the unification scale - can be actually accessible even at the LHC~\cite{Preda:2025afo}.

Regarding dark matter, one loses an inert doublet candidate of the Dirac version of the theory. However, the lightest RH neutrino $N$ can easily play the role of the warm dark matter, with a mass in the keV region, as we discussed in the previous section in the context of the LRSM. If heavy, on the other hand, it can generate the baryon number of the universe through leptogenesis as in the LRSM.  

What remains to be studied is the fate of topological defects in this case, in analogy with the study in~\cite{Senjanovic:2025enc}. We plan to address this issue in near future~\cite{majoranaPS}.

\section{SO(10) theory}
\label{so10}
 
$SO(10)$ grand unified theory~\cite{Fritzsch:1974nn,Georgi:1974my} embeds the Pati-Salam theory in a simple gauge group, and as such, offers a new rich physics of baryon and lepton violation in the form of nucleon decay. It is the first truly unified theory, in a sense that besides unifying strong and electroweak interactions as in $SU(5)$, it also unifies a generation of fermions in a 
 a single $16_{ F}$ dimensional multiplet. Moreover, $16_{ F}$ contains also a right-handed neutrino $\nu^c_L$ (one is forced to work with the fermions of the same chirality for the sake of Lorentz symmetry)
\begin{equation}
    16_{F}^T= \left(q,u^c,d^c,l,e^c,{\color{red}\nu^c}\right)_{ L},
\end{equation}
where in obvious notation $q,l$ denote the quark and lepton weak doublets and $u^c,d^c,e^c$ the anti-quarks and positron weak singlets. 
Evidently, $SO(10)$ provides a natural 
framework~\cite{GellMann:1980vs,Glashow:1979nm} for yet another realization of the seesaw picture, just as the LR symmetric model. I will not discuss it here any further, for a pedagogical review of the subject, see e.g~\cite{Senjanovic:2011zz}. Suffice it to say that in order to generate tree-level mass for the RH neutrino $N$, one needs a large complex $126_H$ complex Higgs scalar representation. This brings in a dillemma: stay renormalisable and use such a huge field, or keep small representations and appeal to higher dimensional operators. Both versions have their pros and cons, as I now briefly discuss. For simplicity, I start with small representations. Even then, there is a choice of the GUT scale Higgs field, since there are basically two equally big representations available: an adjoint, antisymmetric $45_H$ or a symmetric $54_H$ field. The former has been pursued much more, although not clear why. I will comment on both as I go along.

\subsection{A case for higher dimensional operators}
\label{nonrenso10}

If one is willing to give up renormalisability. the situation simplifies considerably - instead of $126_H$ field, one can opt for a small spinorial scalar representation $16_H$. I already commented on the choice of GUT fields, all that is needed is the low energy Higgs scalar containing the usual SM doublet. The natural candidate is clearly the smallest vectorail representation $10_H$. It has to be doubled (or complexified) in order to distinguish between top and bpttom quarks. 

For definiteness, I focus on the adjoint GUT Higgs field, which implies the following set of scalar representations in a minimal such model 
\begin{equation}\label{higgs}
45_{ H}; \,\,\,\,\,  16_{ H}; \,\,\,\,\,  10_{ H}\,,
\end{equation}
in obvious notation. The $45_{ H}$ field, together with the $16_{ H}$ Higgs field, serves to break the GUT down to the Standard Model symmetry, which is then further broken in the usual manner by $10_H$.

\paragraph*{Neutrino mass}\label{numass}

The mass of $N$ is generated by the $16_H$ field whose vev breaks $B-L$ symmetry 
\begin{equation}\label{Nmass}
m_{ N} \simeq \frac {\langle 16_{ H} \rangle^2}{\Lambda}\,,
\end{equation}
where $\Lambda$ denotes the cutoff associated with higher dimensional operators.
There is also two-loop radiatively induced N mass~\cite{Witten:1979nr}, but can be easily shown to be subdominant for any reasonable value of the cutoff~\cite{Preda:2022izo}. 

What happens next is crucial. The third generation of fermions clearly gets mass through the leading $d=4$ terms, and since $\langle 10_H \rangle$ is a singlet under quark-lepton $SU(4)_C$ symmetry, one has $m_{ D_3} = m_{ t}$,
where $m_{ D_3}$ is the third generation neutrino Dirac mass term. From
 $\langle 16_{ H} \rangle = M_{ I}$, using \eqref{Nmass}, one has for neutrino mass
\begin{eqnarray}\label{nueffect}
m_{ \nu} \simeq  \frac {(m_{ D_3})^2} {m_{ N}}  \simeq \frac {m_{ t}^2 \Lambda} {M_{ I}^2}\, .
\end{eqnarray}
In order that neutrino mass be sufficiently small, the intermediate scale has to be almost as large as the GUT scale~\cite{Preda:2022izo} - contradiction with the naive belief of intermediate scales for the sake of unification ~\cite{delAguila:1980qag,Rizzo:1981su}.
Instead. some scalars have to be light in order to ensure gauge coupling unification: a color octet, a neutral weak triplet and an analogue of the quark doublet~\cite{Preda:2022izo}. In order words, the usually assumed desert picture is simply not valid~\cite{Preda:2024vas}, and these light states could in principle even lie at the LHC or the next hadron collider. This nicely contradicts the belief in the so-called extended survival principle, according to which the scalar masses take largest possible values consistent with symmetries in question~\cite{Mohapatra:1982aq}.

\subsection{A case for renormalisability}
~\label{renso10}

If however, one insists on the renormalisability, a minimal model would use the same representations above, with $16_H$ swapped for $126_H$. It makes all the difference in the world. First of all, now neutrino mass puts no constraints on the theory, and RH neutrino mass is basically arbitrary - for all that we know, it could even be light enough to be the dark matter of the universe, as in the LRSM. Second, a proliferation of scalar states would indicate, at least at first glance, a total freedom in the intermediate scales of the theory. In particular, one could ask whether these scales could lie close to their lower experimental limits. This is particularly important for the case of LR symmetry, since $W_R$ could in principle be accessible at the next hadron collider, if not already at the LHC.

\paragraph*{ $W_{ R}$ at colliders?}\label{WR}

 We have recently studied this issue in detail, and found out that, among other results, that the scale of $SU(2)_R$ breaking could be as low as 10 TeV or so~\cite{Preda:2025afo}, tailor made for the present day or near future hadron collider. 
This is rather surprising since it has been a common belief for decades that $W_{R}$-mass should lie at very high energies~\cite{delAguila:1980qag,Rizzo:1981su}, on the order of $10^{11}\,\rm GeV$. That was based on the extended survival principle which is simply not correct in general, as we have seen above.
 Here again, some scalar states violate the survival principle.

\paragraph*{ Renormalisable theory in trouble?}\label{trouble}

However, there is a serious threat to this program, a dark cloud on the horizon. The authors of~\cite{Jarkovska:2021jvw} have performed an admirable study of the consistency of symmetry breaking and unification conditions. Their in-depth analysis, of which we became aware after performing our study, would indicate that the model is not perturbatively viable. There are potential loopholes, such as the lack of the inclusion of fermions in their one-loop study, and so reserve our judgment for the time being. It is however of monumental importance and I for one hope that the issue will be soon settled once for all.

\paragraph*{ What is the renormalisable $SO(10)$ theory?}\label{whatso10}

Notice that we have been calling this model a (not the) minimal renormalisable theory, unlike what most literature does, including~\cite{Jarkovska:2021jvw}. The reason is, as we already said, that the choice of $45_H$ GUT field is by no means dicated by minimality - one can equally choose the symmetric $54_H$ representation. And maybe one actually should do that, since this version of the theory has a beautiful feature of possessing the full Pati-Salam maximal subgroup, unlike in the $45_H$ case, when after the first stage one is left either with LR subgroup without the $SU(4)_C$ symmetry, or QL one that contains only $SU(4)_C$ symmetry, but not the LR one. Moreover, as opposed the $45_H$ case~\cite{Bertolini:2009es}, this model does not suffer from tachynic states at the tree level, which requires a tedious full one-loop study of the complicated Higgs potential. In other words, if the analysis of~\cite{Jarkovska:2021jvw} were to be right, it would actually serve to pave the way for the minimal renormalisable $SO(10)$ theory - it would have to be based on the $54_H$ GUT field. It is worth stressing that this model possesses a beautiful picture of topological defects, such as domain walls bounded by strings, besides magnetic monopoles~\cite{Kibble:1982dd}.


\section{Conclusion and Outlook}
\label{summary}

In this short review, I have tried to convince the reader that discrete symmetries may play a fundamental role in the search for new physics, beyond the SM. I have focused on the LR symmetry, which is naturally taken to be parity, but can as well be charge conjugation. I chose it for a number of reasons. It is among the very first symmetries a child becomes aware of and second, it provides new dynamics in the form of RH gauge currents. Thirdly, it provides a rationale for the parity breaking at low energies, and maybe most important, it led to neutrino mass decades before experiment. Moreover, the minimal LR symmetric model is a full fledged self-contained theory of neutrino mass, which allows us to probe both the origin and the nature of neutrino mass.  

Although originally the LR symmetric theory predicted Dirac neutrino mass, I have here followed a modern day version of the theory which leads to Majorana neutrino mass through the seesaw mechanism. This in turns implies LNV, which besides the text-book low energy neutrinoless double beta decay, also predicts analogous KS process at hadron collider. This is the crux of the theory, and it allows to directly probe the Majorana nature of RH neutrinos. 

I have further discussed the Pati-Salam generalization of the theory which includes quark-lepton unification. The minimal model requires a large scale unification, orders of magnitude beyond experimental accessibility. Besides domain walls as the LRSM, the theory also predicts the existence of magnetic monopoles. The only way out of the cosmological domain wall and monopole problems seems to be a  tiny breaking of LR symmetry, which pushes domain walls out of our horizon and in the process possibly sweeps monopoles away, as suggested originally in the context of the $SU(5)$ grand unified theory. If so, there would remain a clear imprint of gravitational waves, potentially observable in the next generation of searches. 

The LR symmetry finds its most natural role in the context of $SO(10)$ grand unified theory, where it is a gauged discrete symmetry in the form of charge conjugations. This is a truly unified theory, which besides gauge interactions also unifies a family of fermions. The new feature is the nucleon decay, which restricts the unification scale. Moreover, a necessary huge representation at the tree level brings in a plethora new scalar states which open way for low intermediate scales, and a number of low energy processes such as neutron - anti neutron oscillations, for example. One important issue is still not settled, i.e. what constitutes the minimal theory - more precisely what should the GUT breaking Higgs field be. Most of the focus up to now was on the adjoint antisymmetric $45_H$ representation, but there is a possibility of the inconsistencies at the quantum level. If so, one would be forced to use instead the symmetric $54_H$ field which cries for a comprehensive study.

{\bf Acknowledgments}  I wish to thank the organisers of the Ljubljana Discrete2024 symposium for giving me the opportunity to present these ideas here, in spite of not having been able to deliver my talk. Most of the material presented here is based on the work I did originally with Rabi Mohapatra and Wai-Yee Keung, and Vladimir Tello and Michael Zantedeschi in recent years. 

\printbibliography
\end{document}